```
Title: (tn_header.eps)
Creator: Adobe Illustrator(r) 6.0
CreationDate: (17/04/97) (16:46)
```

# Design Patterns for Description-Driven Systems

J.-M. Le Goff, I. Willers

*CERN, Geneva, Switzerland*

R. McClatchey, Z. Kovacs

*Centre for Complex Cooperative Systems, Univ. West of England, Frenchay, Bristol BS16 1QY UK*

*Abstract*

In data modelling, product information has most often been handled separately from process information. The integration of product and process models in a unified data model could provide the means by which information could be shared across an enterprise throughout the system lifecycle from design through to production. Recently attempts have been made to integrate these two separate views of systems through identifying common data models. This paper relates description-driven systems to multi-layer architectures and reveals where existing design patterns facilitate the integration of product and process models and where patterns are missing or where existing patterns require enrichment for this integration. It reports on the construction of a so-called description-driven system which integrates Product Data Management (PDM) and Workflow Management (WfM) data models through a common meta-model.

# 1. Description-Driven Systems and Multi-Layer Architectures

'Description-driven systems' can be defined as systems in which the definition of a domain-specific configuration is captured in a computer-readable form. This definition can be interpreted by applications to achieve domain-specific goals. In a description-driven system definitions are separated and managed independently from instances. This allows the definitions to be specified and to evolve asynchronously from instantiations (and executions) of those definitions. Description-driven systems require computer-readable models both for definitions and for instances. These models are loosely coupled - coupling takes place when instances are created or when a definition, corresponding to existing instantiations, is modified. The coupling is loose since the lifecycle of each instantiation is independent from the lifecycle of its corresponding definition.

One example of the use of description-driven systems is in a workflow management system (WfM [1]) where the business process model defines the instantiated workflows and the definitions are managed separately from the instantiations. WfM systems are often built on a multi-layer architecture [14]. In WfM systems the workflow instances (such as activities or tasks) correspond to the lowest level of system abstraction - the instance layer (see Figure 1). In order to instantiate the workflow objects a workflow scheme is required. This scheme describes the workflow instances and corresponds to the next layer of abstraction - the model layer. In order for the workflow scheme itself to be built, a further model is required to store the semantics for the generation of the workflow scheme. This model (i.e. a model describing another model) is the next layer of system abstraction - the meta-model layer (see Figure 1).

The semantics required to adequately model application-specific information will, in most cases, be different. For example, the semantics for describing Product Data Management (PDM [2]) systems (product types, product composition types etc.) will be very different from those describing WfM systems (activity types, activity composition types, actor types, etc.). To facilitate integration between (in this case PDM and WfM) meta-models a universal type language capable of describing all meta-information is required. The common approach is to define an abstract language which is capable of defining another language for specifying a particular meta-model, in other words meta-meta-information. The accepted conceptual framework for meta-modelling is based on an architecture with four layers: Figure 1 illustrates the four layer meta-modelling architecture adopted by the OMG, based on the ISO 11179 standard.

The meta-meta-model layer is the layer responsible for defining a general modelling language for specifying meta-models. This top layer is the most abstract and must have the capability of modelling any meta-model. It comprises the design artifacts in common to any meta-model. At the next layer down a (domain specific) meta-model is an instance of a meta-meta-model. It is the responsibility of this layer to define a language for specifying models, which is itself defined in terms of the meta-meta types of the meta-meta modelling layer above. Examples of objects at this level from manufacturing include workflow process description, nested subprocess description and product descriptions. A model at layer two is an instance of a meta-model. The primary responsibility of the model layer is to define a language that describes a particular information domain. Example objects for the manufacturing domain would be product, measurement, production schedule, composite product. At the lowest level user objects are an instance of a model and describe a specific information and application domain.

# 2. Features of Description-Driven Systems

Description-driven systems features can be realised through the adoption of a multi-layered architecture. Description-driven systems are flexible and provide many powerful features including reusability, complexity handling, versioning, system evolution and interoperability. This section examines each of these features in turn and explains how a multi-layer architecture facilitates those features.

- *Reusability*. It is a natural consequence of separating definition from instantiation in a system that reusability is promoted. Each definition can be instantiated many times and reused for multiple applications. For example, a single activity definition can be captured in a workflow management system and can be used for many workflow process specifications.

- *Complexity handling (scalability)*. As systems grow in complexity it becomes increasingly necessary to capture descriptions of system elements rather than capturing detail associated with each individual instantiation of an element. Scalability can be eased if descriptive information is held both at the model and meta-model layers of a multi-layer architecture and if information is captured about the mechanism for the instantiation of objects at a particular layer. In a multi-layer architecture there are fewer data and types to manage at each layer but more semantics is needed to cater for system complexity and flexibility. These semantics are provided at the next higher (or descriptive) layer of abstraction. As an example of complexity handling consider the difference between describing the details of every single car of a given model produced by a company and describing the generic details of a model type. Each single instance of a car is derived from a given model type - description should be handled at the type level and details, such as the chassis number, specified only when required for a specific car instance.

- *Version handling*. It is natural for systems to change over time - new elements are specified, existing elements are amended and some are deleted. Element descriptions can also be subject to change over time. Separating description from instantiation allows new versions of elements (or element descriptions) to coexist with older versions that have been previously instantiated. For example, car models change over time and their production processes may need to be revisited as a consequence. Cars of different model versions must be handled over time and coexist with other cars of differing model versions. Separating details of model types from details of single cars allows the model type versions to take place asynchronously with the production of single cars.

- *System evolution*. When descriptions move from one version to the next the underlying system should cater with this evolution. However, existing production management systems, as used in industry, cannot cater for this. In the car example, it is not possible for a single production line to evolve while production is taking place. Rather the production line is flushed of cars following a particular model version before the production line is changed to reflect the requirements of the new model version. Production is therefore not continuous in nature and design changes take time to be rolled forward into production. In capturing description separate from instantiation, using a multi-layer architecture, it is possible for system evolution to be catered for while production is underway and therefore to provide continuity in the production process and for design changes to be reflected quickly into production.

- *Interoperability*. A fundamental requirement in making two distributed systems interoperate is that their software components can communicate and exchange data. In order to interoperate and to adapt to reconfigurations and versions, large scale systems should become 'self describing'. It is desirable for systems to be able to retain knowledge about their dynamic structure and for this knowledge to be available to the rest of the distributed infrastructure through the way that the system is plugged together. This is absolutely critical and necessary for the next generation of distributed systems to be able to cope with size and complexity explosions. A stronger aspect of interoperability is that distributed systems and components to be integrated should have common ways of handling and dealing with system objects such as events, security, systems management, transactions and faults. Software components must be able to plug into these common distributed services and facilities.

## 3. Implementing Description-Driven Systems

Object-oriented systems provide the mechanisms for the capture of system description at a high level of abstraction - descriptive objects themselves have state and methods - and are therefore suitable for building description-driven systems. When implementing a description-driven system based on objects, the descriptive element, which holds information about another object is called a 'meta-object' - meta-objects manage the meta-data required to implement description-driven systems. The 'meta-' prefix is used in the same manner, as it was used for meta-models, i.e. it describes the connection between objects of different layers of abstraction in description-driven systems. Note also that the usage of the term meta-object denotes that the system not only 'stores' the descriptive information but also manages it (i.e. it has data, methods and state).The following sections investigate the design patterns required to integrate process and product model based on a multi-layered architecture.

## 3.1 Patterns

Some design patterns appropriate for handling meta-data in a multi-layer architecture e.g. the Composite and Iterator [3] patterns, have been well-specified. Gamma defines the Composite pattern in the following way: "Compose objects into tree structures to represent part-whole hierarchies. Composite lets clients treat individual objects and compositions of objects uniformly". According to Gamma, the Iterator pattern "provides a way to access the elements of an aggregate object sequentially without exposing its underlying representation". These patterns are studied later in this paper.

Blaha and Premerlani [4], have extended the OMT notation to help specify patterns. A number of patterns have been added to the OMT language to provide "a higher level of building blocks for models than the base primitives of class, association and generalisation". They also introduce cyclicity into the composite pattern. Of particular interest to the subject of this paper are the Graph, Item Description and Homomorphism patterns of [4]. The next sections take patterns described in [4] and enrich these patterns to provide integration of process and product (meta-)models.

## 3.2 Item Description Pattern

Coad's [5] Item Description pattern shows the association between descriptions and instances. In principle this pattern is the manifestation of the relationship between meta-objects and objects. This pattern describes consecutive layers of description-driven systems. The association between Items and Item Descriptions can be an aggregate and support link attributes and qualifiers. In the car example of Section 2, individual cars (of a particular model) are Items which are built according to a single car model description. In other words, the association between car and model holds sufficient semantics for a particular instance of a car to be built according to a model definition. This mechanism is essential to the separation of instantiation from definition, as required by the multi-layer architecture of description-driven systems where semantics are required for the instantiation of Items from a Item Descriptions.

## 3.3 Homomorphism Pattern

Figure 2 also shows the Homomorphism pattern expounded in [6]. This figure shows that two Item Descriptions are themselves related and an association can be defined between them. As a consequence of the Item Description pattern and the fact that semantics have been added to the association between Item Descriptions, there will necessarily be semantics attached to the association of one (instantiated) Item to another (instantiated) Item.

According to [6] "Homomorphisms are most likely to occur for complex applications that deal with meta-data". Since there are relationships between elements in each layer (e.g. a relationship between Item Descriptions) it is natural that the Homomorphism pattern appears between layers. The Homomorphism pattern is therefore fundamental to description-driven systems.

As an example of the use of this pattern, consider two Item Descriptions: the familiar car and car model and a production process and production process model. In this example, there are many instantiations of cars of a particular model and production activities of a particular process model. An association can be specified between a car model and a production process model - i.e. information specific to the execution of a production process model on a specific car model. When an instantiation of the production process is performed on a particular car, details such as the operational conditions must be specified. These operational conditions may be derived from the information on the association between car model and production process model. That is, the semantics of the association between the instantiated Items can be derived from the semantics on the association between the corresponding Item Descriptions.

## 3.4 Version Pattern

In description-driven systems it is important to keep track of versions of definitions and instantiations of these definitions. Figure 3 proposes a Version pattern that can facilitate individual and collective versioning. This pattern provides the functionality of both the CheckIn/CheckOut Model and Composition Models of configuration management (see [7]). In this pattern each VersionedObject manages a set of individual versions of itself, each instance having a versionId and being referred to as a VersionedObjectProperty. Note that, in principle, a VersionedObject and a VersionedObjectProperty make up one object. Properties are separated from attributes in order to distinguish between meta-object data which is either versioned or not versioned respectively. Changing an object's attributes does not version the object whereas changing an object's properties will version it. In addition to handling versioning of an individual object, this pattern allows for versioning of a collected set of objects, called a Release. This is achieved by defining a class of objects called ReleaseManagers which are specialisations of VersionedObjects. ReleaseManagers are versioned and each version, the ReleaseManagerProperty class, manages a collection of versioned objects.

The ReleaseManager maintains a list of added or removed objects in a release. Static versioning is therefore handled by this pattern. The propagation of changes in a release to dependent objects (i.e the notification of changes to dependent objects and the nature of the dependencies) is described in the following section.

## 3.5 Publisher/Subscriber Pattern

To facilitate dynamic version management, which cannot be handled by the Version pattern alone, use can be made of Gamma et al's Observer pattern [3], otherwise referred to as the Publish/Subscribe pattern. Figure 4 shows this pattern. In this pattern a publisher Item (or meta-object) sends out notifications which will reach all subscriber Items without the Publisher knowing who the Subscribers are and how many Subscribers there are. In UML [8] this can be represented by a directed association as shown by the arrow between Publisher and Subscriber in Figure 4. This pattern is useful in handling versions of meta-objects when there are dependencies between the meta-objects but the meta-objects are not tightly coupled.

## 3.6 Graph Pattern

The general form of a graph is shown in Figure 5 where nodes are linked to other nodes. In an undirected graph an edge connects any two nodes, whereas in a directed graph an edge connects a source node to a sink node. In addition, a directed graph can have nodes with any number of edges. Complex graphs make a distinction between branch and leaf nodes, whereas simple graphs do not. The example quoted by [4] to describe complex directed graphs is that of a simple Unix file structure: files are either data files or directory files and a directory file contains named files which are identified by a filename that is unique in the context of a directory file. In the Unix file system a file can belong to multiple directories via symbolic links and a file may have a different name in each directory where it is referenced - this means the structure is a graph. All files have a parent directory except the root file as shown in Figure 6 (from [4]). The graph is complex since distinction is drawn between datafiles and directory files - datafiles being leaf nodes and directory files being branch nodes.

In an acyclic graph, when the graph has been traversed repetitively from parent to child nodes, there are no instances where traversal leads to a node being a child of itself. Cyclic graphs can allow this form of recursion. Therefore the complex directed graph of the Unix file system example is acyclic in nature, since a directory file cannot contain a reference to itself at any level. The complex Directed Acyclic Graph pattern of [4] does not allow semantics to be added to the association between nodes as branches (see Figure 5), consequently there is no way of identifying, and associating attributes or methods to, a particular instance of the link.

## 4. Providing Description-Driven System Features Using Meta-Objects

### 4.1 Handling Complexity

In Section 2, it was stated that scalability can be eased, if descriptive information is held both at the model and meta-model layers of a multi-layer architecture. The Item Description pattern combined with the Directed Acyclic Graph pattern provides the mechanism by which this can be achieved.

Figure 7 shows a combination of the Item Description and Directed Acyclic Graph patterns. The combination of the patterns is established by decomposing an Item Description into its constituent Item Descriptions. In other words, an Item Description can be either elementary or composite in nature and therefore some Item Descriptions can be made up of other Item Descriptions. Consider the car and car model example of earlier. A particular description of a car model is composed of other descriptions: e.g descriptions of the engine, the chassis, the drive-system (front-axle system, rear-axle system, wheels, tyres etc.). Some of these descriptions are elementary e.g wheels and some composite e.g drive-system. The association between a Composite Item Description and its children will hold semantics such as the number of constituent descriptions of a common type (e.g 4 wheels of 1 wheel description). However, as stated earlier, a simple combination of the Item Description and Directed Acyclic Graph patterns as described in [4] does not enable the identification of a particular constituent Item Description within its Composite Item Description. In the example, it is not possible to determine which wheel is located at which wheel position.

Consequently the combined patterns require enrichment by the introduction of another meta-object which captures the membership of an Item Description within its Composite Item Description(s). Figure 8 shows the Enriched Directed Acyclic Graph and Item Description combined pattern. An Item Description can be part of many different Composite Item Descriptions e.g. one wheel description could be employed in both the front-axle system and the rear-axle system. One instance of the Composite Member meta-object will hold the full semantics of the membership of a particular Item Description in a single Composite Item Description. In other words, it is possible to determine which wheel is located in which axle system and in which location in that axle system. When a particular Item Description is instantiated into an Item the composition of that Item is determined by traversing the graph of its Item Description. The result will be a hierarchy of Items organised as a tree in which each node is of a particular Item Description. In the car example the car is made up of a chassis, an engine, a drive-system (comprising front- and rear-axle systems each of which is composed of 2 wheels etc.). The tree is as deep as there are layers in the directed acyclic graph and each composite node will have a number of constituent nodes equal to the number of Composite Member meta-objects in the Item Description corresponding to that node (see Figure 9).

The complexity of the overall model of Items is therefore handled through the reuse of Item Descriptions. The reuse can take place at any point in the traversal of the directed graph as long as the graph is acyclic. For the car example, the number of Items and the number of levels of compositeness is not great and complexity handling is not a major issue. As either the complexity of the Item and the number of levels of composition increases the role of Composite Member meta-objects becomes essential. The Enriched Directed Acyclic Graph and Item Description patterns have been used to manage the complexity inherent in the construction of a large scientific apparatus, as described in Section 5 of this paper.

## 4.2 Integrating Product & Process Models

Having discussed the role of a directed acyclic graph to describe Items and their constituents in the previous section, it is now proposed that any product or any process can be modelled in terms of an Enriched Directed Acyclic Graph pattern combined with an Item Description pattern.

In manufacturing, models are used to support the design life cycle of a particular product [9]. Products can evolve over time, their designs may change or the production process may be improved. Earlier it was stated that PDM systems have been employed to manage product data in the design life cycle. PDM systems traditionally employ hierarchies to capture product composition (so-called 'Bill Of Materials', BOM) and therefore, as the complexity of the product grows PDM systems suffer from a products explosion. Basing a PDM model on an Enriched Directed Acyclic Graph pattern combined with an Item Description pattern, handles the products explosion. The consequence is that the BOM is only available once the product composition tree has been generated by traversal of the complete graph structure (as shown in Figure 9).

Products are subject to many processes in the manufacturing life cycle such as design processes, assembly processes, test processes, maintenance processes etc. Each of these processes can be complex and composite in nature. Ideally the description of these processes should be captured in a model and instances of these processes managed in some repository. One example of a process management system is a WfM, which, as stated earlier, can be described as description-driven systems. Process models to support systems such as WfMs must cope with process composition, process sequence, parallelism of processes and synchronisation of processes. Basing a WfM model on an Enriched Directed Acyclic Graph patterns combined with an Item Description pattern supports processes of arbitrary complexity including composition and sub-process reuse. Furthermore, using the CompositeMember meta-objects of the Enriched Directed Acyclic Graph pattern allows the capture of process sequence, parallelism and synchronisation.

There is an increasing movement in manufacturing to integrate product and process models for the purposes of life cycle data management. Therefore any system which can manage both product and process information in a common model is very desirable to the manufacturing community. Basing both a product and a process model on the above patterns and associating product descriptions with process descriptions, provides a uniform model for manufacturing. The association between the two descriptions carries semantics in that it describes how a particular process description is applied to a particular product description and any conditions or constraints on how the process acts on the product. This association of process description to product description is very powerful - it allows different associations to be defined between a product and different processes that can take place throughout its life cycle e.g. design, assembly, testing, maintenance etc. For example, the association of a maintenance process to a product will require quite different conditions to be captured from those that are captured when a design process is carried out on that same product. The integration of PDM with WfM demonstrates the power of a unified product and process life cycle model.

## 4.3 Handling Evolution

Production systems should cater for the evolution of product or process descriptions regardless of the current state of the production and even while the production continues. Since, as discussed earlier, layers in a description-driven system are only loosely coupled, modifications in the meta-model layer can be carried out asynchronously from the application of those modifications in the model layer (which is itself defined in and generated from the meta-model layer). Similarly, modifications in the model layer can be asynchronously applied from their instantiations in the instance layer.

Even though the modifications are asynchronously applied in each layer, notification of the modification is required to provide traceability in the production systems. This mechanism can be handled through a combination of the Publish/Subscribe pattern, described in Section 3.5, the Item Description pattern of Section 3.2 and the Version pattern of Section 3.4. In the combination of these patterns, an Item Description is a concrete Publisher and any Item associated with this Item Description is a concrete Subscriber. A modification in the Item Description (at the model layer) is then notified to its Subscribers (at the instance layer) which can apply their modifications when appropriate.

The application of the Subscribers' modifications follows the Homomorphism pattern, described in Section 3.3. The Homomorphism pattern provides linkage between versions of Items and Item Descriptions. Consequently an Item can determine the consequences to itself of moving to a new version of an instantiation of its Item Description.

## 4.4 Meta-Objects and Standardisation

In distributed object-based systems, object request brokers provide for the exchange of simple data types and, in addition, provide location and access services. The CORBA standard is meant to standardise how systems interoperate. OMG's CORBA Services [10] specify how distributed objects should participate and provide services such as naming, persistent storage, life cycle, transaction, relationship and query. The CORBA Services standard is an example of how self describing software components can interact to provide interoperable systems.

Recently a considerable amount of interest has been generated in meta-models and meta-object description languages. Work has been completed within the OMG on the Meta Object Facility [11] (MOF) which is expected to manage all kinds of meta-models relevant to the OMG Architecture. The purpose of the OMG MOF is to provide a set of CORBA interfaces that can be used to define and manipulate a set of interoperable meta models. The MOF uses CORBA interfaces for creating, deleting, manipulating meta objects and for exchanging meta models.This meta-modelling approach will facilitate further integration between product data management and workflow management thereby providing consistency between design and production and speeding up the process of implementing design changes in a production system. The MOF provides the mechanisms required to bring together OMG work on Product Data Enablers [12] and the Workflow Management Facility [13], [14].

The usage of the MOF will depend very much on viewpoint. From a systems designers viewpoint, who will be looking down the layers of a multi-layer architecture, the MOF is used to define an information model for a particular domain of interest. Another viewpoint is that of a systems programmer who is looking up the multi-layer architecture. CORBA clients use the MOF to obtain information model descriptions which support reflection and interoperability.

## 5. The CRISTAL Project

A prototype has been developed which has facilitated a study of description-driven systems. The objective of this prototype was to integrate a Product Data Management (PDM) model with a Workflow Management (WfM) model in the context of the CRISTAL (Cooperating Repositories and Information System for Tracking Assembly Lifecycles) project currently being undertaken at CERN, the European Centre for Particle Physics in Geneva, Switzerland.

The Compact Muon Solenoid (CMS) experiment [15] currently being constructed at the European Centre for Particle Physics at CERN will comprise several complex detectors for fundamental particle physics research. Each detector is being constructed out of, potentially, over a million parts and will be produced and assembled during the next decade by specialised centres distributed world-wide (see Figure 10). Each constituent part of each detector must be accurately measured and tested locally prior to its ultimate assembly at CERN. The CRISTAL system [16], [17] is being developed to control the production and assembly process of the CMS Electromagnetic Calorimeter (ECAL) detector. It employs workflow (WfMS) and product data management (PDM) techniques to provide an infrastructure in which the engineering data can be warehoused.

A distributed object-oriented database (Objectivity) is used to hold both the engineering data and the definitions of the detector components and of the tasks which are performed on the components. The ECAL construction database follows an object-oriented design to maximise flexibility and reusability. An approach has been taken, which promotes self-description and a degree of data independence. In adopting such a description-driven design approach, a separation of object instances from object descriptions instances was needed. This abstraction resulted in the delivery of a meta-model as well as a model for CRISTAL.

A CRISTAL system comprises one or more distributed data gathering centres (Figure 10), each of which is federated into the system. These centres are a single Central System and multiple Local Centres in which the CRISTAL software will run. Each Local Centre will have a set of measurement Instruments defined in the database in terms of the commands that each instrument uses and the data formats expected as outcomes from the execution of workflow activities by instruments [17].

Figure 11 shows the software architecture of a Local Centre. The software comprises a set of Instrument Agents, a set of Product Managers for handling all part data to/from the database, a Local Centre Manager which supervises the data gathering in a centre, a set of Digital Control Panels (DCPs) which handle user interaction with CRISTAL and a Data Duplication Manager (DDM) which handles all duplication of data between the Local Centre and the Central System.

Product Managers (PM) provide the mechanism by which products (or detector parts) are tracked through workflows. They manage concurrency of workflow activity executions, they manage data storage and carry out all book-keeping of events. Any significant occurrence that happens to a product in a workflow is recorded by its PM which effectively keeps track of the 'state' of the product in production. As the PM handles all database activity for specific products, the DCPs are protected from any changes occurring in the database schema. As a consequence, the PM lies at the heart of the CRISTAL system and controls access to versions of product and process data.

The PM uses the present state of products and workflow activities with the prevailing production conditions to determine the next step that a product takes in its workflow. In effect the Product Manager acts like a mediator (as defined in the Mediator pattern of [3]) liaising with both workflow activity and product objects, which, themselves, have no knowledge about each other's state. By using the Mediator pattern, workflow activity objects are decoupled from product objects thereby allowing interaction with these objects to be handled separately through a single object. By so doing the protocol for interaction with both workflow activity and product objects is simplified and only implemented through the mediator object.

In practice, this means that the product (PDM) and process (WfM) information can be managed separately but can be combined by the Mediator process. Product and process data are stored as database objects, managed by the PM. All events associated with these product and process objects are also stored in chronological order as database objects. The PM interrogates the database for product and process 'state' information and combines this with the production conditions to determine the next viable workflow activities that can be performed on the product under consideration.

Not only does the PM manage static product and process information, it also handles evolving product and process information. When a new release of the product specification becomes available at a centre, the PM will compute when and how the change can be applied. Changes can be of different types (e.g product, workflow activity, data format) and as a mediator the PM will forward the handling of the change to the appropriate database object. As a consequence of its role in determining when a change can be applied, it is possible for products of the same definition to be following quite different versions of the production scheme at any one time in a centre.

It is the responsibility of each products's PM to keep track of the product in its workflow. Products can move between centres, so PMs in both the source and destination centres must be able to follow the same workflow i.e to suspend and resume workflow activities when products are shipped between centres. The Product Manager has been implemented as a CORBA object using the CORBA lifecycle service. In its role as manager of the Local Centre the LCM manages the life cycle of each PM instance, including when a product is physically transported from one centre to another ('shipped').

Figure 12 shows an enriched Homomorphism pattern replacing the condition element of Figure 2 by a mediator class. The PM carries out the physical role of mediation in the CRISTAL prototype. This enriched pattern describes both the relationship between the meta-model and model layers of CRISTAL, through the familiar Item Description pattern and the role of the PM as the mediator between Items of the model layer (in this case between products and workflow activities).

## 6. Conclusions

The following existing patterns emerged from the CRISTAL data model: Item Description, Publish/Subscribe, Homomorphism, Graph, Tree and Mediator. These patterns were shown to be insufficient to provide the flexibility required in the CRISTAL data model. However, with enrichment of the Graph, Tree and Homomorphism patterns and the addition of a new Version pattern it has been possible to provide the functionality required to integrate PDM with WfM, using a description-driven approach.

Figure 13 summarises the inter-relationship between the set of patterns identified for the integration of PDM systems and WfM systems in the CRISTAL data model. It shows a Versioned Graph Pattern which has been derived from the Version, Complex Graph and Publish/Subscribe patterns. It also shows an Enriched Homomorphism pattern which has been derived from the Item Description and Mediator patterns. Furthermore the diagram brings together the Complex Tree pattern with the Versioned Graph pattern from which it is derived. Instantiation of the Complex Tree pattern from the Versioned Graph pattern is performed by the Product Manager which consequently carries out the integration of the PDM and WfM aspects and acts as the CRISTAL execution component.

Blaha and Premerlani [4] state that "patterns provide a higher level of building blocks for models than the base primitives of class, association and generalisation". This work has shown that this assertion is not only true for models but can be extended to include meta-models.

The concept of using meta-data to reduce complexity and aid navigability of data resident in a database is well known. Also its use in minimising the effect of schema evolution in object databases has been stated many times elsewhere. In the CRISTAL project meta-data are used for these purposes and, in addition, meta-models are used to provide self-description for data and to provide the mechanisms necessary for developing a query facility to navigate multiple data models.

Foote and Yoder [18] have applied the concepts of pattern representations to the domain of data description. They conclude that candidate patterns are required to describe meta-data structures and their inter-relationships. Design patterns are thus needed in object-oriented design to describe meta-schemae such as CRISTAL meta-objects.

The meta-model approach to design reduces system complexity, provides model flexibility and can integrate multiple, potentially heterogeneous, databases into the enterprise-wide data warehouse. A first prototype for CRISTAL based on CORBA, Java and Objectivity technologies has been deployed in the autumn of 1998 [19]. The second phase of research will culminate in the delivery of a production system in 1999 supporting meta-model queries and the definition, capture and extraction of data according to physicist-defined viewpoints.

## Acknowledgments

The authors take this opportunity to acknowledge the support of their institutes and in particular thank P. Lecoq, J-L. Faure, M. Pimia and J-P Vialle. A. Bazan, F. Estrella, T. Le Flour, C. Koch, S. Lieunard, S. Murray, G. Organtini, L. Varga, M. Zsenei and G. Chevenier are thanked for their assistance in developing the CRISTAL prototype.

**Figures**

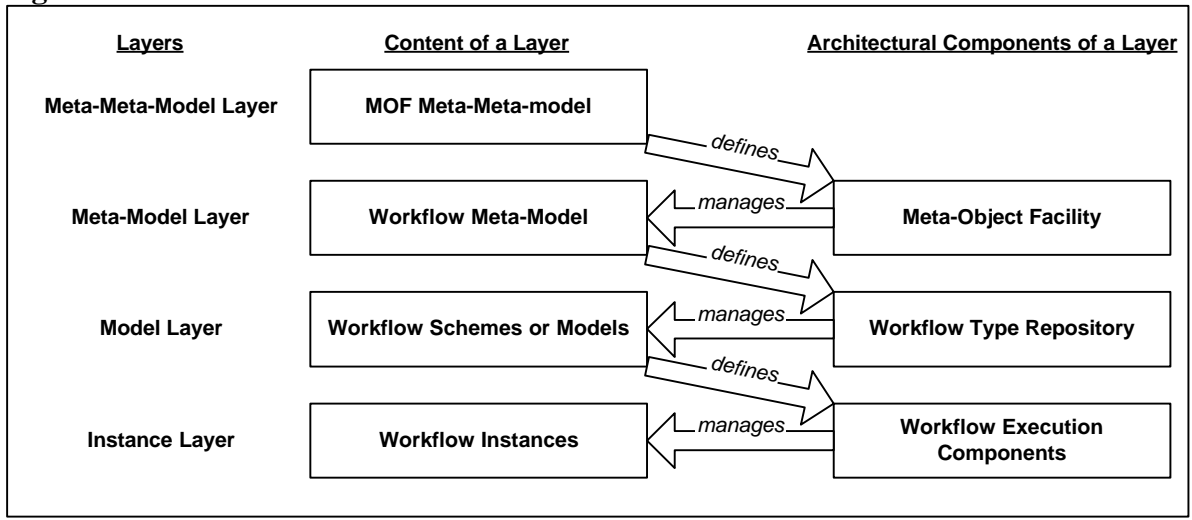

Figure 1: A four-layer meta-modelling architecture.

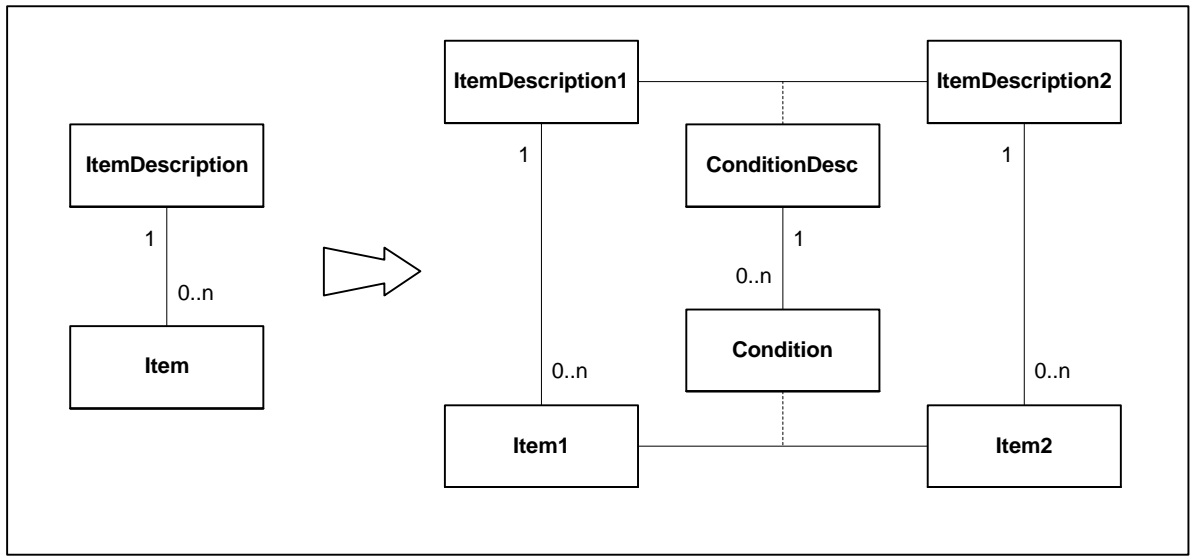

Figure 2: Item Description and Homomorphism patterns.

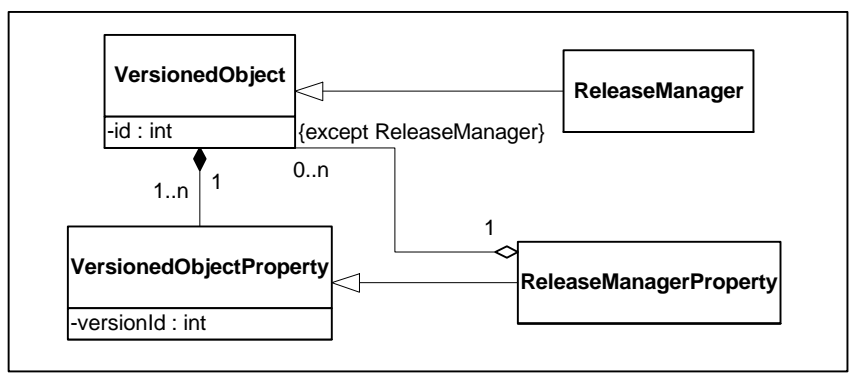

Figure 3: Version pattern.

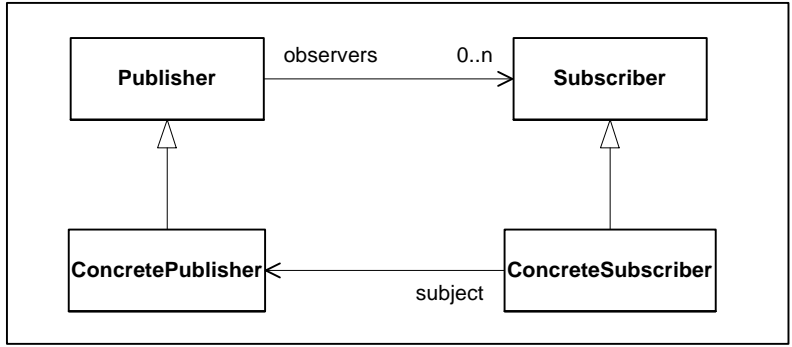

Figure 4: Publisher/Subscriber pattern.

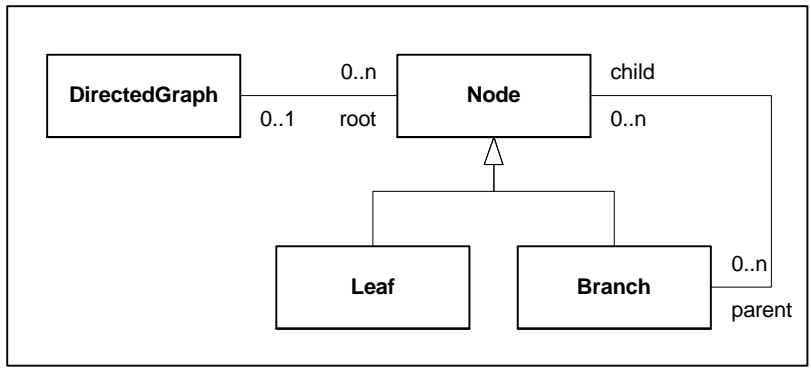

Figure 5: Complex Directed Graph pattern.

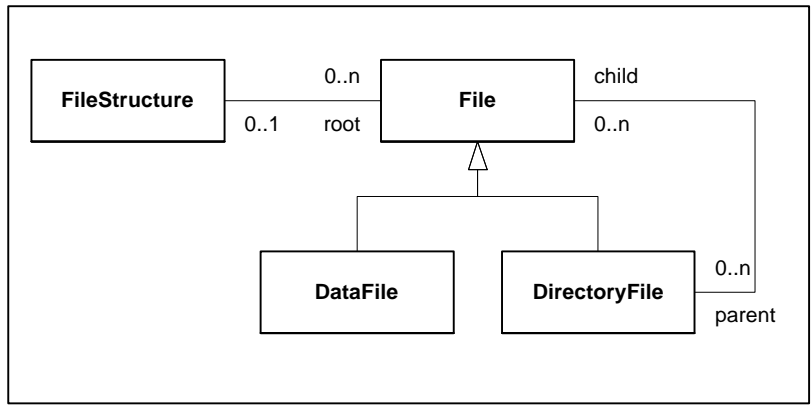

Figure 6: File directory as an example Graph pattern.

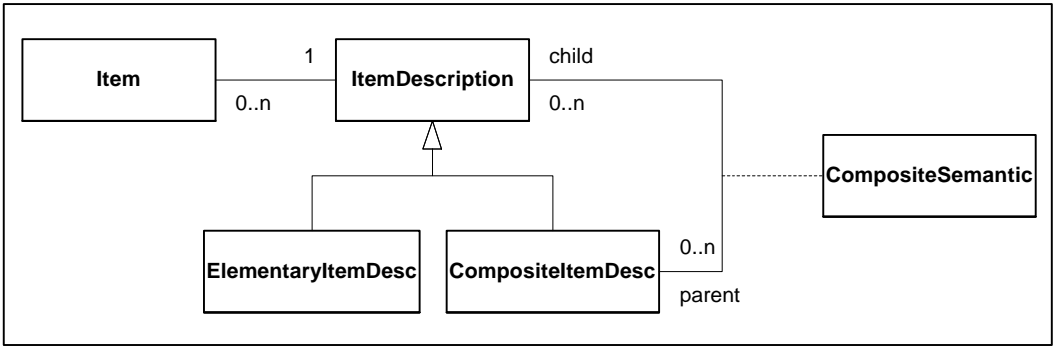

Figure 7: Combination of the Item Description and Directed Acyclic Graph patterns.

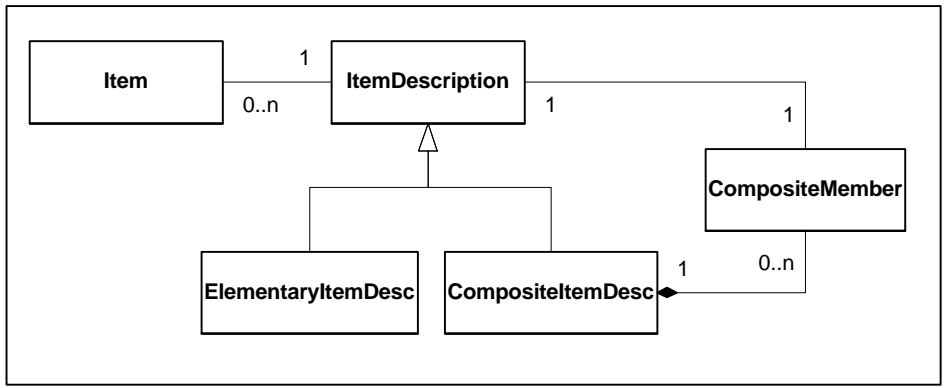

Figure 8: Enriched Directed Acyclic Graph pattern

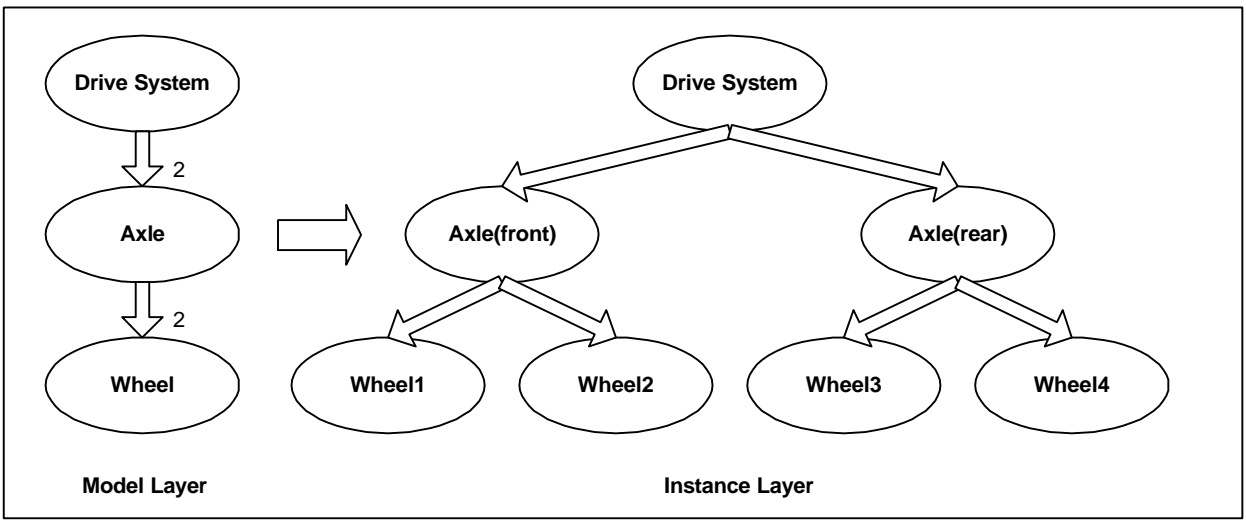

Figure 9: Car example for Graph handling complexity.

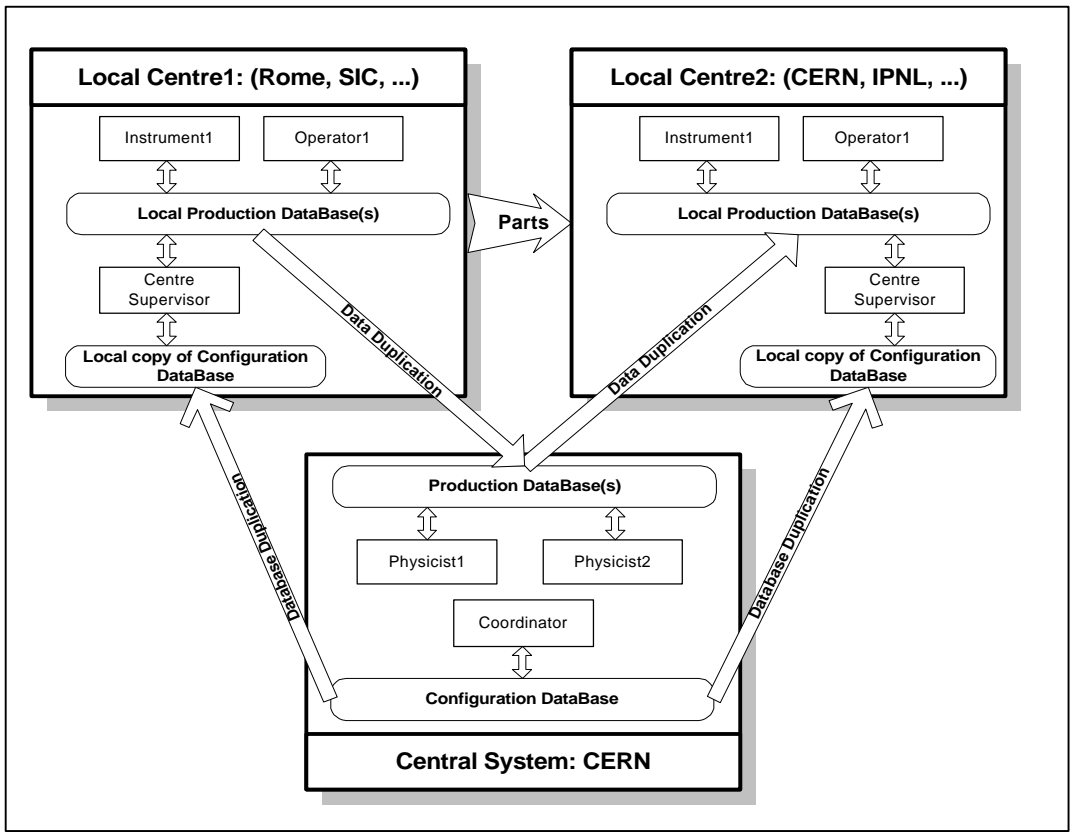

Figure 10: The CRISTAL Multi-Centre architecture.

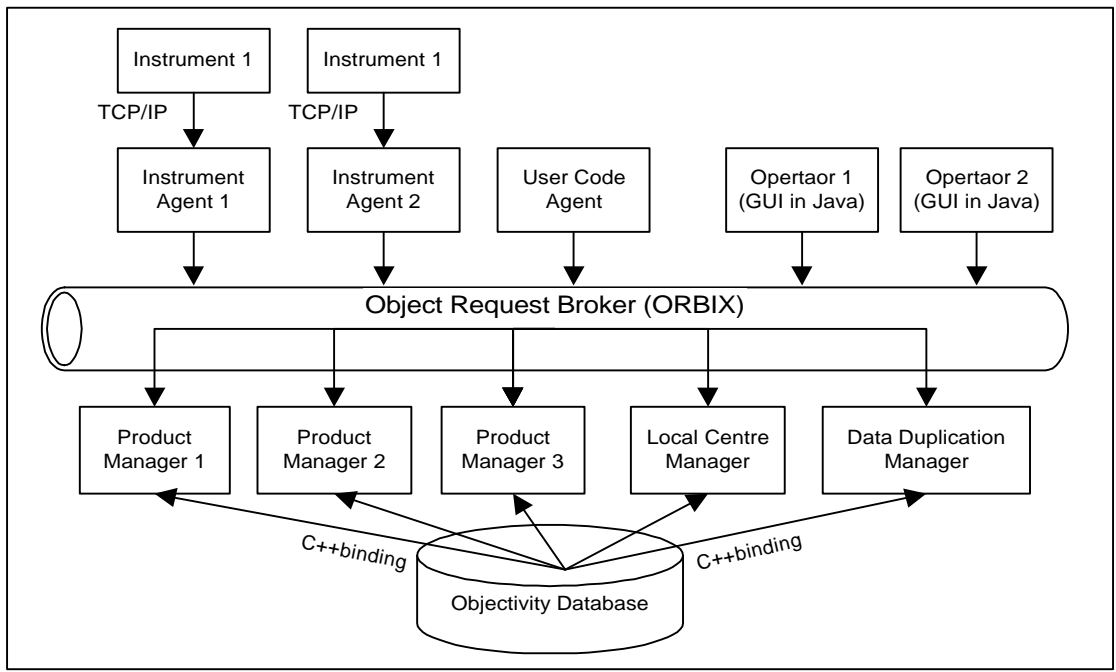

Figure 11: Local Centre software architecture.

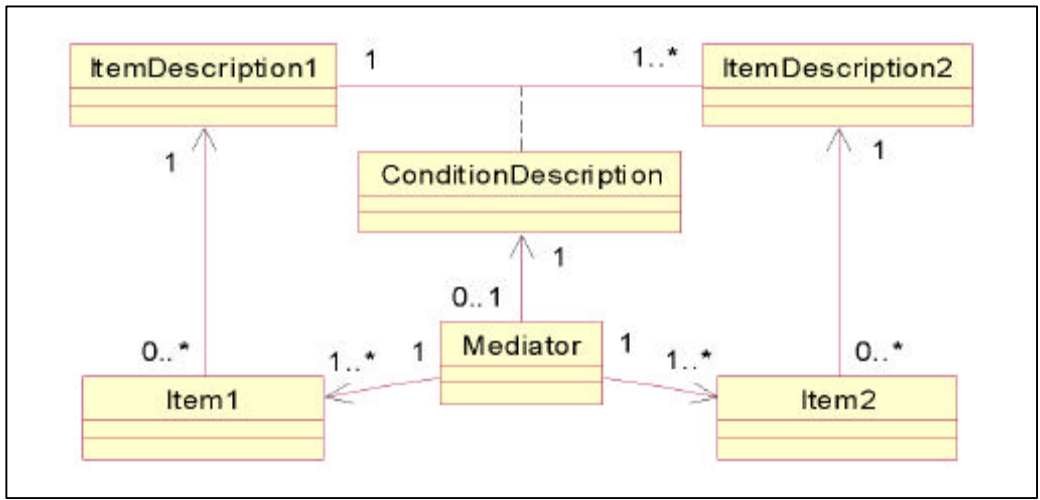

Figure 12: Enriched Homomorphism pattern.

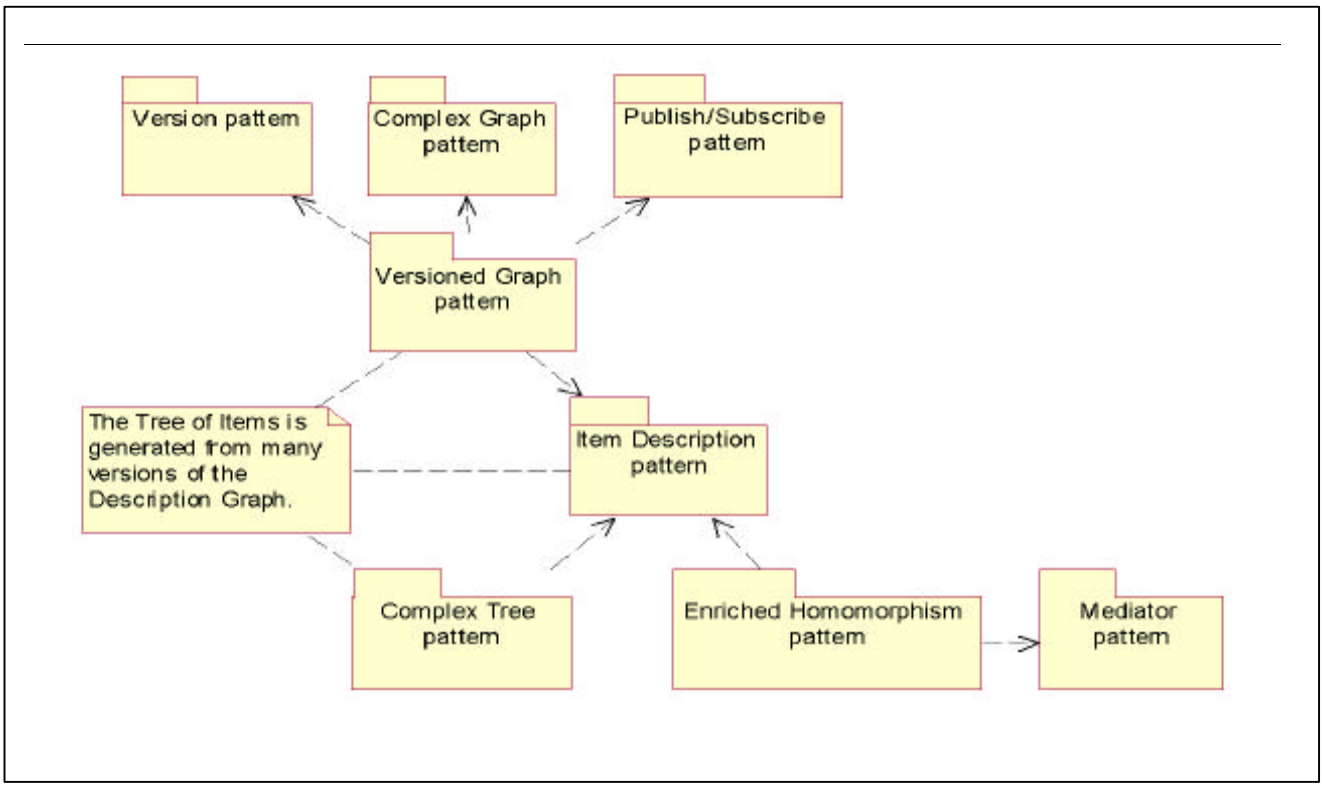

Figure 13: CRISTAL pattern summary.